\documentclass[%
 reprint,
 amsmath,amssymb,
 aps,
]{revtex4-2}
\usepackage{braket}
\usepackage{xcolor}
\usepackage{color}
\usepackage{graphicx}
\usepackage{dcolumn}
\usepackage{bm}
\usepackage{bbm}
\usepackage[export]{adjustbox}
\usepackage[colorlinks=true,linktocpage=true,breaklinks=true]{hyperref}

\begin{document}

\preprint{APS/123-QED}


\title{Influence of the surface states on the nonlinear Hall effect in Weyl semimetals}
\author{Diego Garc\'ia Ovalle}
\email{email: diego-fernando.garcia-ovalle@univ-amu.fr}
 \affiliation{Aix-Marseille Université, CNRS, CINaM, Marseille, France.}
 
 \author{Armando Pezo}
 \email{email: armando-arquimedes.pezo-lopez@univ-amu.fr}
 \affiliation{Aix-Marseille Université, CNRS, CINaM, Marseille, France.}

 \author{Aurélien Manchon}%
 \email{email: aurelien.manchon@univ-amu.fr}
 \affiliation{Aix-Marseille Université, CNRS, CINaM, Marseille, France.}

\date{\today}
\begin{abstract}
 We investigate the influence of surface states on the nonlinear Hall response of non-centrosymmetric time-reversal invariant Weyl semimetals. To do so, we perform a tomography of the Berry curvature dipole in a slab system using a minimal two-band model. We find that in the type-I phase, the nonlinear Hall response is not particularly sensitive to the presence of Fermi arcs or other trivial surface states. However, in the type-II phase, we find that these surface states contribute substantially to the Berry curvature dipole, leading to a strong thickness dependence of the nonlinear Hall response. This feature depends on the nature of the surface states and, henceforth, on the slab geometry adopted. In order to assess the validity of this scenario for realistic systems, we performed Berry curvature dipole calculations by first principles on the WTe$_2$, confirming the dramatic impact of surface states for selected slab geometries. Our results suggest that surface states, being topological or not, can contribute much more efficiently to the nonlinear Hall response than bulk states. This prediction is not limited to topological semimetals and should apply to topologically trivial non-centrosymmetric materials and heterostructures, paving the way to interfacial engineering of the nonlinear Hall effect.
\end{abstract}

\maketitle


\section{Introduction}

Conventional wisdom inherited from Hall's foundational work \cite{Hall1879,Hall1881} states that Hall currents flowing transverse to the injected charge direction are only permitted as long as time-reversal symmetry is globally broken. In other words, ordinary, anomalous or topological Hall effects only exist either in the presence of an external magnetic field or in magnetic materials displaying a net magnetization \cite{Nagaosa2010}. However, this long-lived statement has been recently challenged by two important observations. First, it has been realized that time-reversal does not need to be broken globally, but only locally. In fact, as long as no crystal symmetry reinstalls time-reversal effectively, Hall currents are allowed. This is particularly true for certain classes of antiferromagnets with a non-collinear magnetic configuration \cite{Chen2014,Kubler2014,Nakatsuji2015,Nayak2016} (see also Ref. \cite{Bonbien2022}). Second, the time-reversal symmetry breaking necessary to obtain Hall effect is not required anymore at the second order in the electric field. Under certain conditions, anomalous Hall effect can appear in non-magnetic materials to the second order of the electric field \cite{Sodemann,Ma,Kang}. Whereas the anomalous Hall effect in ferromagnetic and antiferromagnetic compounds is associated with the Berry curvature of the material's ground state, the nonlinear anomalous Hall effect (NLHE) is rather associated with the Berry curvature dipole (BCD). From a symmetry standpoint, the minimal requirement is inversion symmetry breaking which ensures that the Berry curvature does not vanish, but this is not sufficient: mirror symmetry also needs to be broken to obtain a finite BCD. Sodemann and Fu \cite{Sodemann} identified the crystallographic point groups that possess the minimal requirements for the observation of NLHE, a study recently refined by Du et al. \cite{Du2021} to include both intrinsic (BCD related) and extrinsic mechanisms allowed by symmetry.

Different material candidates have been explored experimentally and theoretically as suitable options to obtain NLHE. From an experimental point of view, quadratic responses in the electric field have been detected, among others, in bilayers and few layers of WTe$_{2}$ \cite{Ma,Kang} which have $C_{2v}$ point group. From a theoretical perspective, other possibilities for large BCD values have been proposed, including transition metal dichalcogenides \cite{Du,Jhih,He,Singh,YZhangTMD,Zhou}, graphene \cite{Battilomo,Pantaleon,ZhangTBG}, and especially Dirac \cite{Samal} and Weyl semimetals (WSM) \cite{Zeng}. Ab initio simulations \cite{YZhang} have stimulated further theoretical studies of BCD in 3D WSMs, because of their rich geometrical features and their potential benefits to create highly efficient electronic transport devices. NLHE in WSMs has also been verified analytically by applying perturbation theory \cite{FZhang}, suggesting that the transport is sensitive to intraband processes, the chemical potential and the tilting of the Weyl nodes. Besides, similar optical effects such as second harmonic generation can be confirmed by applying Floquet \cite{Morimoto} and many-body quantum formalisms \cite{Rostami}. Last but not least, a full Green function theory of the NLHE has been proposed recently \cite{Du2021}, pointing out the differences between extrinsic and intrinsic contributions. In this article, we are interested on the nonlinear response arising on the intrinsic mechanism driven by the BCD.

Nonmagnetic WSMs such as TaAs or WTe$_2$ are particularly interesting platforms for the realization of NLHE because inversion symmetry is necessarily broken and Berry curvature diverges at the Weyl nodes. WSMs possess pairs of doubly degenerate linearly dispersive states, forming Weyl cones at Fermi level \cite{Wan2011}. According to the Nielsen-Ninomiya theorem \cite{Nielsen}, each pair of nodes carries Berry curvature monopoles of opposite chirality which are connected via Fermi arcs lying at opposite surfaces of the slab \cite{Armitage}. Type I WSMs, such as elemental Tellurium \cite{DRodriguez}, Janus superlattices \cite{Mengt1} and Ta or As compounds \cite{BQLv,SYXu,YanTaAs}, are characterized by point-like Fermi surface in the bulk and vanishing density of states. Type II WSMs, such as MoTe$_2$ \cite{DengMoTe2,Tamai} and WTe$_2$ \cite{Bruno,Wu} but also the magnetic candidate Co$_3$Sn$_2$S$_2$ \cite{Guin,Morali,Li,Wang2018,Okamura2020,Ikeda2021}, offer a slightly different paradigm as the Weyl cone spectrum is tilted in momentum space, breaking Lorentz invariance. As a result, the Weyl points arise at the boundary between electron and hole pockets. Notice that certain compounds can support type I as well as type II Weyl nodes \cite{Meng,MZhang}. \par

A remarkable aspect of WSMs is the nature of their surface states. As mentioned above, alike topological insulators WSMs possess topologically protected surface states in the form of spin-momentum locked Fermi arcs that connect bulk Weyl nodes of opposite chirality. In type I WSMs, the Fermi arcs coexist with the projection of electron (or hole) pockets when the chemical potential lies away from the neutrality point. In type II WSMs, the Fermi arcs coexist with projected electron and hole pockets irrespective of the value of the chemical potential, as well as with trivial closed loops called "track states" \cite{McCormick}. As a consequence, surface states of WSMs can be rich, resulting in enhanced Edelstein effect \cite{Johansson2018}, and unconventional patterns in quantum oscillation experiments \cite{Potter2014,Bulmash2016,Wang2016c} (see also Ref. \cite{Moll2016}). Previous works pointed out that topological materials defined in slab geometries can exhibit interesting transport properties due to finite size effects and the behavior of surface states inside the samples. In this context, it has been shown that the anomalous Hall conductivity is highly influenced by surface states such as Fermi arcs in Weyl systems without time-reversal symmetry \cite{Breitkreiz}, even in presence of disorder \cite{ChestaSlab}. Additional studies have been performed in confined geometries to clarify, among others, the behavior of chiral magnetic effects \cite{Gorbar}, the magnetoresistance \cite{Alekseev} and the quantum Hall effect in Dirac semimetals \cite{Schumann}.

\begin{figure}[ht!]
\includegraphics[width=1.1\linewidth]{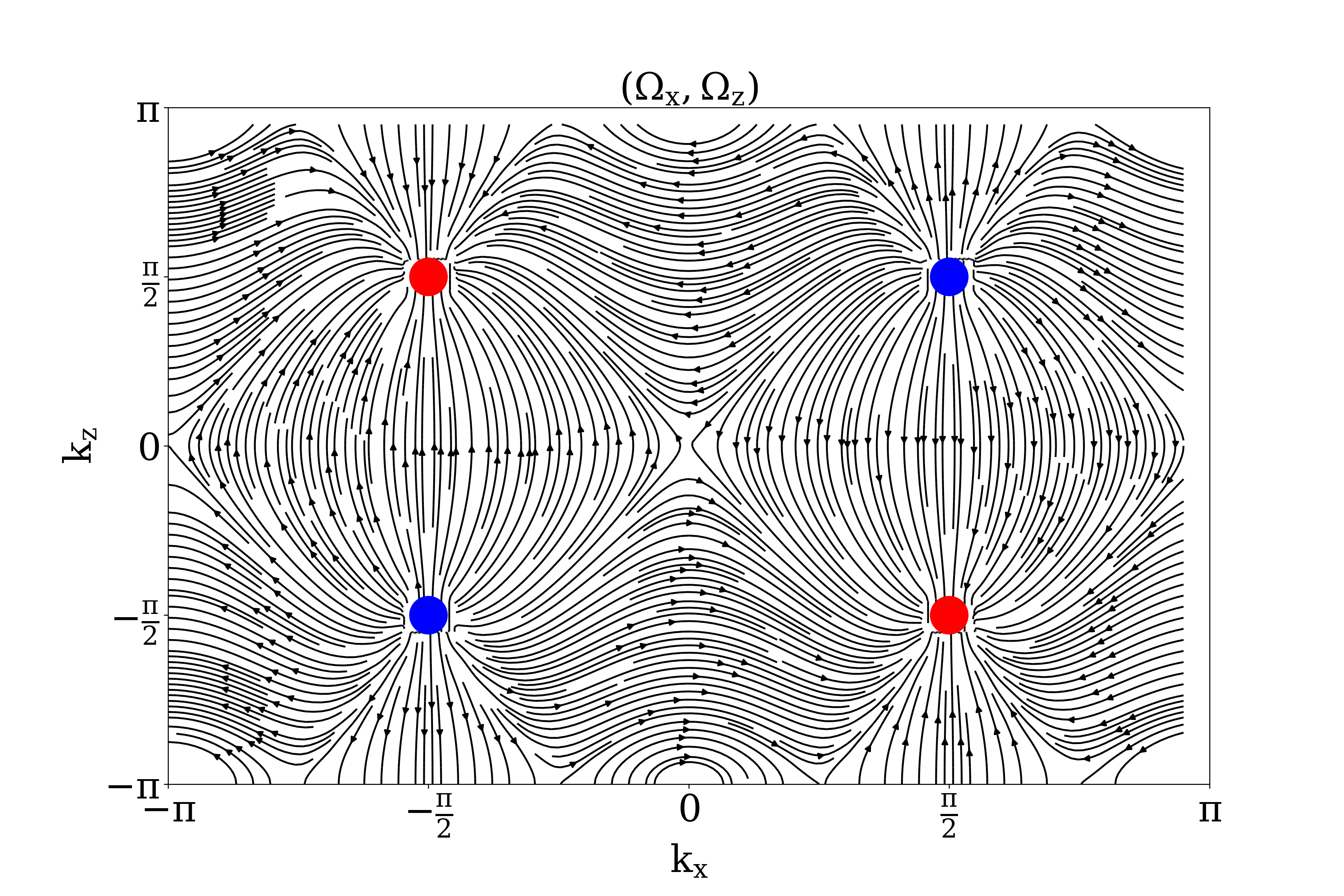}
\caption{(Color online) Berry curvature of the Hamiltonian \eqref{4} in the ($k_x,\;k_z$) plane with $k_y=0$ and intrinsic parameters $k_0=\frac{\pi}{2}$, $m=2$, $t_x=\frac{1}{2}$, $\gamma=1$ and $t=1$. Note that the Berry curvature of the two-band model is not sensitive to the value of $\gamma$ \cite{McCormick}. The solid lines represent the Berry curvature vector field $(\Omega_x,\;\Omega_z)$, whereas the blue (red) dots represent the local positive (negative) divergencies taking place at the Weyl  nodes.} 
\label{FigureBC}
\end{figure}

In this work, we seek to understand how the surface states of nonmagnetic WSMs influence the NHLE response driven by the BCD. To do so, we consider a minimal 2-band model of a time-reversal invariant WSM with inversion symmetry breaking in a slab geometry, so that bulk and surface states are treated on equal footing \cite{McCormick}. This model exhibits four Weyl points: the minimum number of degeneracies due to time-reversal symmetry. These four points are associated with local divergencies of the Berry curvature, as depicted on Fig. \ref{FigureBC}. In this sense, after neglecting the vanishing components of the BCD tensor due to mirror symmetries in the 3D lattice, Zeng et al. \cite{Zeng} recently reported that NLHE requires Weyl cone tilting and an asymmetric Fermi surface when the nodes lie at the same energy. The NLHE is also influenced by the distance between nodes. Accordingly, our study gives further insight about the implications of the Fermi arc configurations on the BCD. Remarkably, it also complements a recent study that comprises a surface BCD due to the projection of Fermi arcs in type I WSMs \cite{Wawrzik}.

Our paper is organized as follows. In Section \ref{II}, we recall the general formalism for nonlinear Hall transport driven by BCD and basic physical considerations for our model of interest. In Section \ref{III}, we provide and discuss our results regarding the connection between surface states and NLHE, its dependency on the Weyl cones tilting and its layer decomposition. In Section \ref{IV}, we also collate these outcomes with realistic numerical simulations on a WTe$_2$ slab with different cuts. Finally, we summarize and state our main conclusions in Section \ref{V}.

\section{General Theory and  Model} \label{II}

Let us begin by recalling basic elements of NLHE theory \cite{Sodemann}. We start with a nonmagnetic crystal and its corresponding Bloch Hamiltonian $\mathcal{H}_{\mathbf{k}}$, whose eigenstates $\ket{u_{n\mathbf{k}}}$ satisfy Schr\"odinger equation $\mathcal{H}_{\mathbf{k}}\ket{u_{n\mathbf{k}}}=\epsilon_{n\mathbf{k}}\ket{u_{n\mathbf{k}}}$ with eigenenergies $\epsilon_{n\mathbf{k}}$. If the crystal is subjected to a sufficiently small electric field $\mathbf{E}=$Re$\left[\mathbf{\mathcal{E}}\exp(i\omega t)\right]$, $\mathbf{\mathcal{E}}\in \mathbb{C}^3$, such that the adiabatic limit is still valid, and assuming weak disorder, the first order Hall current is induced by the Berry curvature \cite{Nagaosa2010}

\begin{align}
    \mathbf{\Omega}_{n\mathbf{k}}=i\sum_{m\not=n} \frac{\braket{u_{n\mathbf{k}}|\mathbf{\mathbf{\hat{v}}_{\mathbf{k}}}|u_{m\mathbf{k}}}\times \braket{u_{m\mathbf{k}}|\mathbf{\mathbf{\hat{v}}_{\mathbf{k}}}|u_{n\mathbf{k}}}}{(\epsilon_{n\mathbf{k}}-\epsilon_{m\mathbf{k}})^2},\label{1}
\end{align}

\noindent 
where $\mathbf{\hat{v}}_{\mathbf{k}}=\partial_\mathbf{k}\mathcal{H}_{\mathbf{k}}$ is the velocity operator. Upon time reversal operation $\mathcal{T}$ we have $\mathcal{T}\mathbf{\Omega}_{n\mathbf{k}}\mathcal{T}^{-1}=-\mathbf{\Omega}_{n-\mathbf{k}}$, and thus the anomalous Hall effect vanishes. Nonetheless, when inversion symmetry $\mathcal{P}$ is further broken, then $\mathcal{P}\mathbf{\Omega}_{n\mathbf{k}}\mathcal{P}^{-1}\not=\mathbf{\Omega}_{n-\mathbf{k}}$, and one can show that to the lowest order in scattering time $\tau$, the (rectified) second order Hall current reads \cite{Sodemann}

\begin{align}
    j_a^0= \left(\frac{e^3\tau}{2\hbar^2}\right) \epsilon_{adc}D_{bd}\mathcal{E}_b\mathcal{E}_c^*.\label{2}
\end{align}
\noindent 
Here, latin indices refer to the components of the usual cartesian basis and $\epsilon_{adc}$ is the Levi-Civita tensor. $D_{bd}$ is the BCD defined as \cite{Sodemann}

\begin{align}
    D_{bd}=\int_{BZ} \frac{d^3{\bf k}}{(2\pi)^3} \left[\sum_n (\mathbf{v}_{n\mathbf{k}})_b(\mathbf{\Omega}_{n\mathbf{k}})_d\frac{\partial f_{n\mathbf{k}}}{\partial \epsilon_{n\mathbf{k}}}\right],\label{3}
\end{align}
with $\mathbf{v}_{n\mathbf{k}}=\partial_\mathbf{k}\epsilon_{n\mathbf{k}}$ the eigenvalues of the velocity operator and $f_{n\mathbf{k}}$ the Fermi distribution function. In the zero-temperature limit, $\partial_{\epsilon_{n\mathbf{k}}}f_{n\mathbf{k}}\to -\delta(\epsilon_{n\mathbf{k}}-\mu)$, setting $\mu$ as the chemical potential. Equation \eqref{2} neglects the intrinsic contribution to NLHE that appears when time-reversal is broken as well as higher order extrinsic contributions (i.e., side-jump and skew scattering) \cite{Du2021}.

The nonmagnetic WSM slab is built on the spinless 2-band model in a cubic lattice introduced in Ref. \cite{McCormick}. The bulk Hamiltonian reads

\begin{align}
    &\mathcal{H}_B=\gamma(\cos2k_z-\cos k_0)(\cos k_x-\cos k_0)\hat{\sigma}_0\nonumber\\
    &-\left[m(1-\cos^2k_x-\cos k_y)+2t_x(\cos k_z-\cos k_0)\right]\hat{\sigma}_1\nonumber\\
    &-2t\sin k_y\hat{\sigma}_2-2t\cos k_x\hat{\sigma}_3 \label{4},
\end{align}

\noindent 
with $\hat{\sigma}_i$, $i=1...3$ being the $2\times2$ Pauli matrices, $\hat{\sigma}_0=\mathbbm{1}_{2\times 2}$ and the cubic Brillouin zone is $\mathcal{C}=\left[-\pi,\pi\right]^3$. The parameters of Eq. \eqref{4} are set to $k_0=\frac{\pi}{2}$, $m=2$, $t_x=\frac{1}{2}$ and $t=1$. This WSM possesses four Weyl nodes of zero energy located at $\mathbf{k}^*=\pm \frac{\pi}{2}(\hat{x}+\hat{z})$ (see Fig. \ref{FigureBC}). Importantly, for the parameters adopted in this work, $\gamma=2$ sets the transition point between type I ($\gamma<2$) and type II ($\gamma>2$) WSM phases. In the remaining of this paper, the transport properties will be investigated as a function of the chemical potential $\mu$ as well as of the tilting of the Weyl cones controlled by $\gamma$. 
Interestingly, Eq. \eqref{4} is constrained by the following mirror symmetries \cite{Zeng}

\begin{align}
    \mathcal{M}_x^{\dag}\mathcal{H}_B(k_x,k_y,k_z)\mathcal{M}_x&=\mathcal{H}_B(-k_x,k_y,k_z),\label{5}\\
     \mathcal{M}_z^{\dag}\mathcal{H}_B(k_x,k_y,k_z)\mathcal{M}_z&=\mathcal{H}_B(k_x,k_y,-k_z),\label{6}
\end{align}
\noindent 
which impose that the Fermi arcs connecting the Weyl nodes lie on the ($z,\;x$) surfaces. In addition, these symmetries constrain the BCD tensor, so that the only non-vanishing elements are $D_{zx}$ and $D_{xz}$. In other words, a second order Hall current can only be obtained in the ($y,\;z$) ($\equiv D_{zx}$) and ($x,\;y$) ($\equiv D_{xz}$) planes. Consequently, the Fermi arcs cannot contribute to the NLHE response. Nevertheless, as discussed below, other trivial surface states can substantially impact the second order response.
 
 We now design the slabs by discretizing Eq. \eqref{4} along a given direction, $\hat{x}$, $\hat{y}$ or $\hat{z}$, of the cubic Brillouin Zone (see, e.g., Ref. \cite{Aurelien}). The Hamiltonian loses periodicity along the chosen axis. The new Hamiltonian $\mathcal{H}_S^{\hat{n}}$ of size $2L\times 2L$, with $\hat{n}=\hat{x},\;\hat{y},\;\hat{z}$ depending on the growth direction, and $L$ the number of layers, is given by
 
\begin{align}
\mathcal{H}_S^{\hat{n}}=\left(
\begin{array}
[c]{cccccc}%
\mathcal{H}_0^{\hat{n}} & \mathcal{H}_1^{\hat{n}} & \mathcal{H}_2^{\hat{n}} & 0 & ...& 0 \\
\mathcal{H}_1^{\hat{n}\dag} & \mathcal{H}_0^{\hat{n}} &\mathcal{H}_1^{\hat{n}} &\mathcal{H}_2^{\hat{n}} & \ddots &\vdots\\
\mathcal{H}_2^{\hat{n}\dag} & \mathcal{H}_1^{\hat{n}\dag} & \ddots & \ddots & \ddots & 0\\
0 & \mathcal{H}_2^{\hat{n}\dag}  & \ddots& \ddots & \ddots & \mathcal{H}_2^{\hat{n}}\\
\vdots & \ddots & \ddots & \ddots & \ddots & \mathcal{H}_1^{\hat{n}}\\
0 & \hdots & 0 & \mathcal{H}_2^{\hat{n}\dag} & \mathcal{H}_1^{\hat{n}\dag} & \mathcal{H}_0^{\hat{n}}
\end{array}
\right).\label{7}
\end{align}

For the block matrices in Eq. \eqref{7}, $\mathcal{H}_0^{\hat{n}}$ is the intralayer Hamiltonian that retains in-plane periodicity after the cut, while $\mathcal{H}_1^{\hat{n}}$ and $\mathcal{H}_2^{\hat{n}}$ are the nearest neighbor and second nearest neighbor interlayer Hamiltonian, respectively. For the cut along $\hat{x}$, the block matrices are given by 

\begin{align}
    \mathcal{H}_0^{\hat{x}}&=-\gamma\cos k_0(\cos2k_z-\cos k_0)\hat{\sigma}_0\nonumber\\
    &-\left[m\left(\frac{1}{2}-\cos k_y\right)+2t_x\left(\cos k_z-\cos k_0\right)\right]\hat{\sigma}_1\nonumber\\
    &-2t\sin k_y\hat{\sigma}_2,\label{8}\\
    \mathcal{H}_1^{\hat{x}}&=\frac{\gamma}{2}\left(\cos 2k_z-\cos k_0\right)\hat{\sigma}_0-t\hat{\sigma}_3, \label{9}\\
     \mathcal{H}_2^{\hat{x}}&=\frac{m}{4}\hat{\sigma}_1. \label{10}
\end{align}

A cut along $\hat{y}$ leads to 

\begin{align}
    \mathcal{H}_0^{\hat{y}}&=\gamma(\cos2k_z-\cos k_0)(\cos k_x-\cos k_0)\hat{\sigma}_0\nonumber\\
    &-\left[m\sin^2k_x+2t_x(\cos k_z-\cos k_0)\right]\hat{\sigma}_1\nonumber\\
    &-2t\cos k_x\hat{\sigma}_3,\label{11}\\
    \mathcal{H}_1^{\hat{y}}&=\frac{m}{2}\hat{\sigma}_1+it\hat{\sigma}_2.\label{12}
\end{align}

Finally, a cut along $\hat{z}$ gives

\begin{align}
     \mathcal{H}_0^{\hat{z}}&=-\gamma\cos k_0(\cos k_x-\cos k_0)\hat{\sigma}_0\nonumber\\
    &-\left[m(1-\cos^2k_x-\cos k_y)-2t_x\cos k_0\right]\hat{\sigma}_1\nonumber\\
    &-2t\sin k_y\hat{\sigma}_2-2t\cos k_x\hat{\sigma}_3,\label{13}\\
    \mathcal{H}_1^{\hat{z}}&=-t_x\hat{\sigma}_1,\label{14}\\
    \mathcal{H}_2^{\hat{z}}&=\frac{\gamma}{2}\left(\cos k_x-\cos k_0\right)\hat{\sigma}_0.\label{15}
\end{align}

\begin{figure}[ht!]
\includegraphics[width=9.95cm,center]{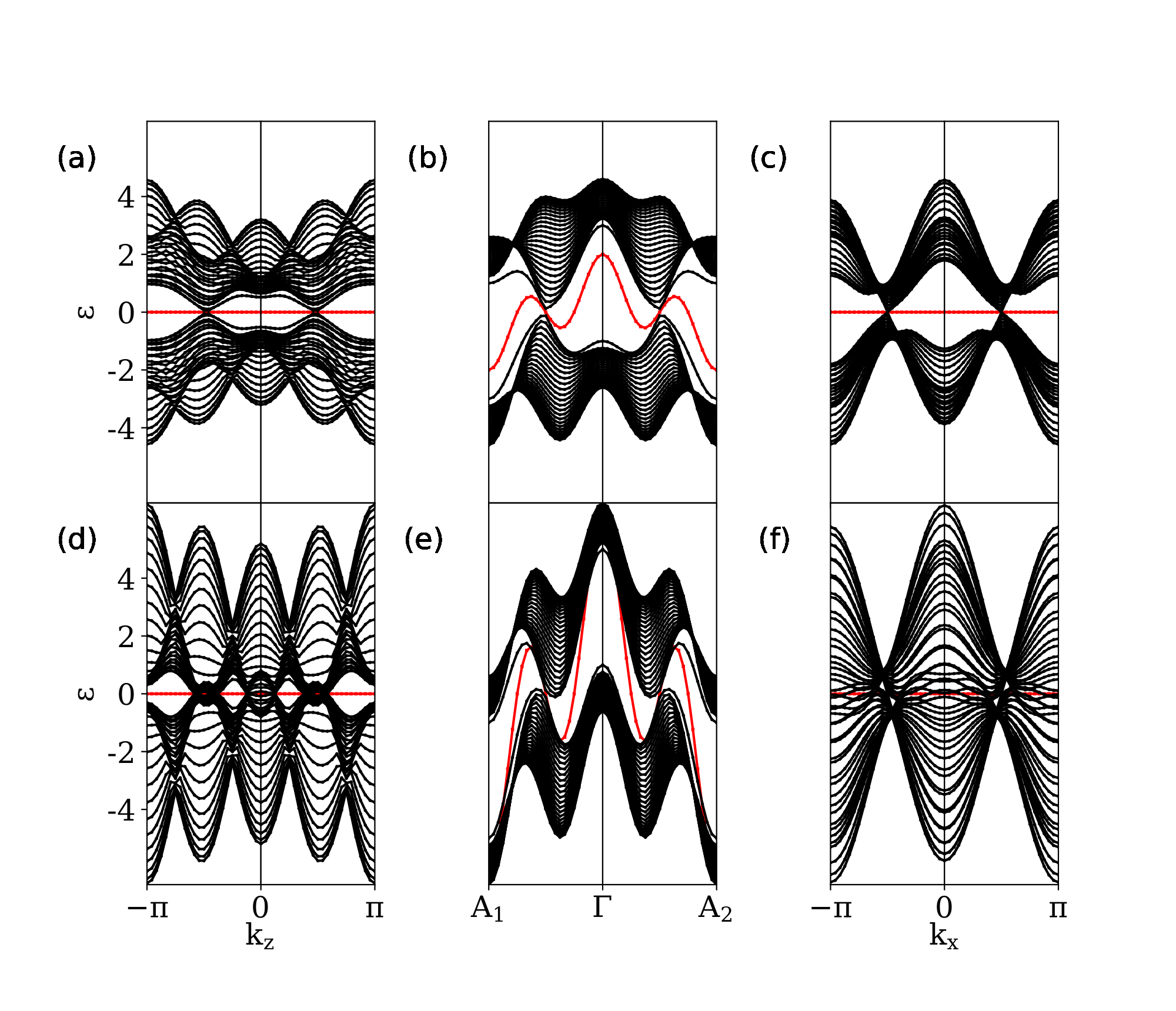}
\caption{(Color online) Band structures of the WSM slab of 25 layers, with $\gamma=1$ (a-c) and $\gamma=3$ (d-f) along the high symmetry path of the cubic lattice for different cuts: (left panels) $\hat{x}$ for $k_y=0$ and $k_z\in [-\pi,\pi]$, (center) $\hat{y}$ along the path $A_1=(-\pi,0,-\pi)\to \Gamma =(0,0,0)\to A_2=(\pi,0,\pi)$ and (right) $\hat{z}$ for $k_y=0$ and $k_x\in [-\pi,\pi]$. The central panels clearly displays type I (b) and type II (e) Weyl cones. The surface states at the top and bottom surfaces are represented with a solid red line. The flat red bands in (a), (c), (d) and (f) are surface states at zero energy that are highly localized at the corresponding surfaces.}
\label{Figureband}
\end{figure}

We illustrate the slab band structures for each cut in Fig. \ref{Figureband}, for $\gamma=1$ (type I WSM) and $\gamma=3$ (type II WSM). Since the slab Hamiltonians retain a Berry curvature perpendicular to the plane of the slab, we define $D_{zx}^{\hat{x}}$ as the BCD associated with $\mathcal{H}_S^{\hat{x}}$ when the electric field is applied along $\hat{z}$, and $D_{xz}^{\hat{z}}$ is the BCD associated with $\mathcal{H}_S^{\hat{z}}$ when the electric field is along $\hat{x}$. These are the only relevant BCD components for the slab geometries presented above due to the symmetry restrictions. We now move forward to the method for calculating the layer decomposition of transport coefficients in the slabs. Since the periodic part of the Bloch function can be written in terms of the complete layer basis as $\ket{u_{n\mathbf{k}}}=\sum_{l=1}^L\ket{\omega_{l\mathbf{k}}}\braket{\omega_{l\mathbf{k}}|u_{n\mathbf{k}}}$, one can extract information about layer $l$ by applying the projection operator $\mathcal{S}_l=\ket{\omega_{l\mathbf{k}}}\bra{\omega_{l\mathbf{k}}}$, $l\in [1,L]$ to an observable $\mathcal{O}$. In this way, it is easy to see that $\mathcal{O} = \sum_{l=1}^L \mathcal{S}_l\mathcal{O}$ and $\sum_{l=1}^L\mathcal{S}_l=\mathbbm{1}_{2L\times 2L}$. We use this description to separate the contributions of the density of states for a slab with a fixed number of layers,

\begin{align}
    \mathcal{D}_{l\mathbf{k}}&=-\frac{1}{\pi}\operatorname{Im}\left[\operatorname{Tr}\left(\mathcal{S}_lG^R_{\mathbf{k}}\right)\right],\label{16}
\end{align}

\noindent 
with the Green's function

\begin{align}
  G^R_{\mathbf{k}}&=\lim_{\eta\to 0^+}\left[(\epsilon+i\eta)\mathbbm{1}-\mathcal{H}_S^{\hat{n}}\right]^{-1}.\label{17}
\end{align}

\noindent 
This decomposition provides an insight about the influence of the surface states for each growth direction. Additionally, in order to obtain the contribution of layer $l$ to the NLHE response along, say, $\hat{y}$, one simply needs to perform the substitution $\hat{v}_{y}\to \mathcal{S}_l \hat{v}_{y}$ into Eq. \eqref{1} and then into \eqref{3}.

\section{Results and Discussion} \label{III}
\subsection{Band structure and surface states}
\begin{figure}[ht!]
\includegraphics[width=0.9\linewidth]{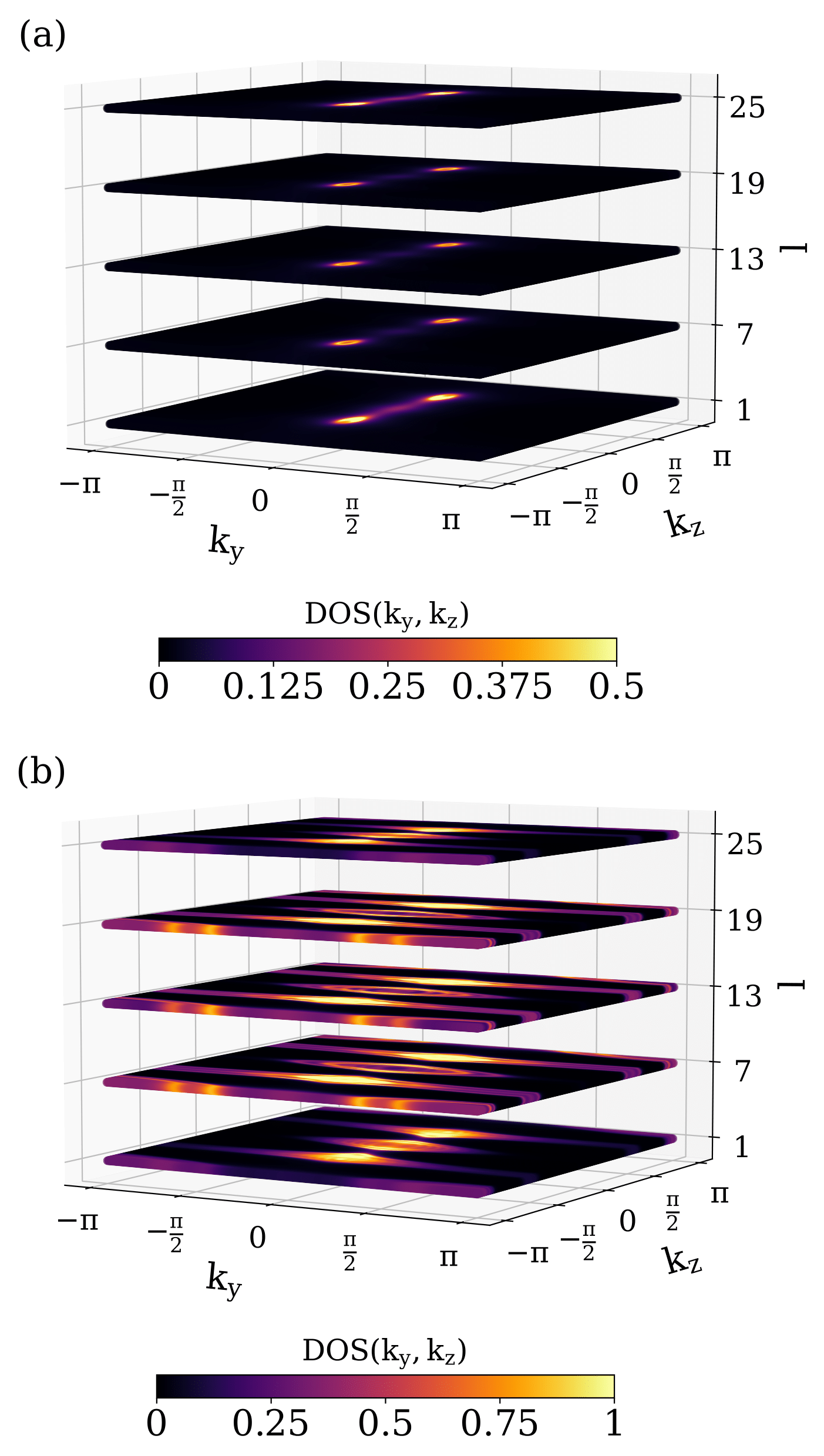}
\caption{(Color online) Projected density of states (DOS) for $\hat{n}=\hat{x}$ as a function of the momentum coordinates for selected layers $l$ for the type I ($\gamma=1$) (a) and type II ($\gamma=3$) phase (b). In the type I phase, the Weyl nodes dominate in the bulk and get connected by degenerate Fermi arcs at the surface, whereas in the type II phase, large trivial Fermi pockets appear across the slab.}
\label{Figuredosx}
\end{figure}

\begin{figure}[ht!]
\includegraphics[width=0.9\linewidth]{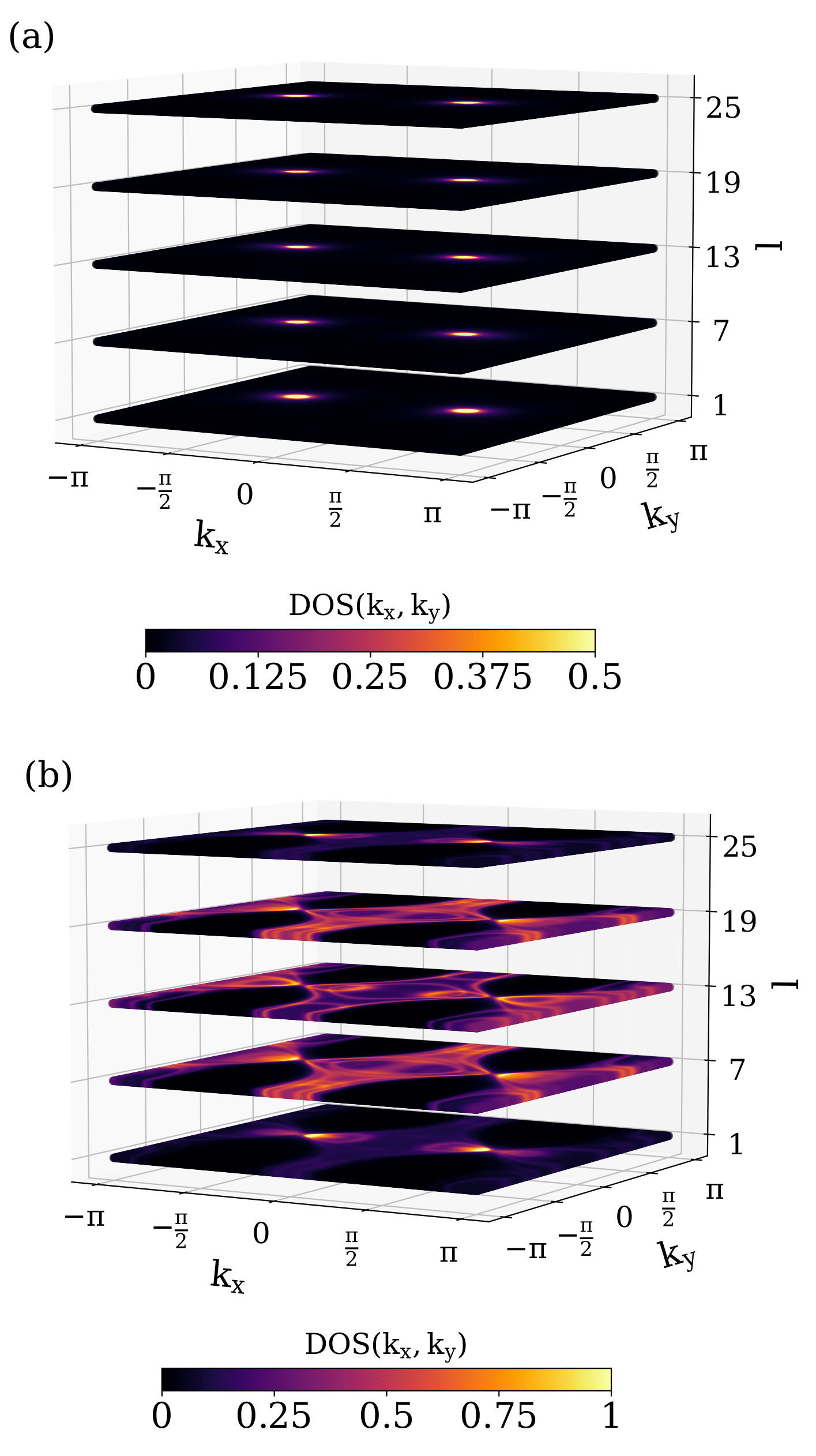}
\caption{(Color online) Projected density of states (DOS) for $\hat{n}=\hat{z}$ as a function of the momentum coordinates for selected layers $l$ for the type I ($\gamma=1$) (a) and type II ($\gamma=3$) phase (b). In the type I phase, the Weyl nodes remain disconnected across the slab, whereas in the type II phase, trivial Fermi pockets dominate in the bulk with track states and projected Weyl nodes at the surface.}
\label{Figuredosz}
\end{figure}

Let us first consider how the Weyl point tilting impacts the slab band structure and its surface states. Since only $D_{zx}^{\hat{x}}$ and $D_{xz}^{\hat{z}}$ are non-vanishing, and because the Weyl nodes are located in the $(k_x,\;k_z)$ plane, we focus on slabs composed of $L=25$ layers and normal to the $\hat{x}$ and $\hat{z}$ directions. Without loss of generality, we fix $\epsilon=0.2$ and select the exemplary cases $\gamma=1,\;3$ to illustrate the differences between a type I and a type II WSM, respectively. To better understand how bulk and surface states evolve upon tilting the Weyl cones, we represent the density of states in the 2D Brillouin zone and projected on a set of representative layers, ranging from the bottommost surface to the topmost surface. Our results are depicted in Fig. \ref{Figuredosx} for the $\hat{x}$-cut and Fig. \ref{Figuredosz} for the $\hat{z}$-cut.  \bigskip 

Figures \ref{Figuredosx} and \ref{Figuredosz} show different behaviors of the bulk and surface states when tuning the tilting parameter. When $\hat{n}=\hat{x}$ (Fig. \ref{Figuredosx}), we notice two distinct situations depending on the value of $\gamma$: if $\gamma=1$ [type I, Fig. \ref{Figuredosx}(a)], the density of states is dominated by the Weyl nodes across the whole slab, with a surface state composed of the projected nodes connected by degenerate Fermi arcs. When $\gamma=3$, [type II, Fig. \ref{Figuredosx}(b)], the density of states of the central layer is composed of electron and hole pockets touching at the Weyl nodes, as expected. These pockets result in large trivial surface Fermi pockets enclosing the nodes, whose connections change direction from $\hat{z}$ (type I) to $\hat{x}$ (type II) WSMs \cite{McCormick}. 

On the other hand, for $\hat{n}=\hat{z}$ (Fig. \ref{Figuredosz}), we obtain a different behavior. Whereas for $\gamma=1$ [type I, Fig. \ref{Figuredosz}(a)], the Weyl nodes remain disconnected throughout the slab, for $\gamma=3$ [type II, Fig. \ref{Figuredosz}(b)], a surface contribution appears due the emergence of track states \cite{McCormick}. Nonetheless, the surface Fermi pockets associated with the projection of the bulk electron and hole pockets remain very small and the surface states are dominated by the Weyl nodes, in sharp contrast with the $\hat{x}$-cut discussed in Fig. \ref{Figuredosx}. We therefore expect the surface states to have completely different impact on the NLHE signal in these two different slab geometries.

\subsection{Nonlinear Hall response}
\begin{figure}[ht!]
\includegraphics[width=1.02\linewidth]{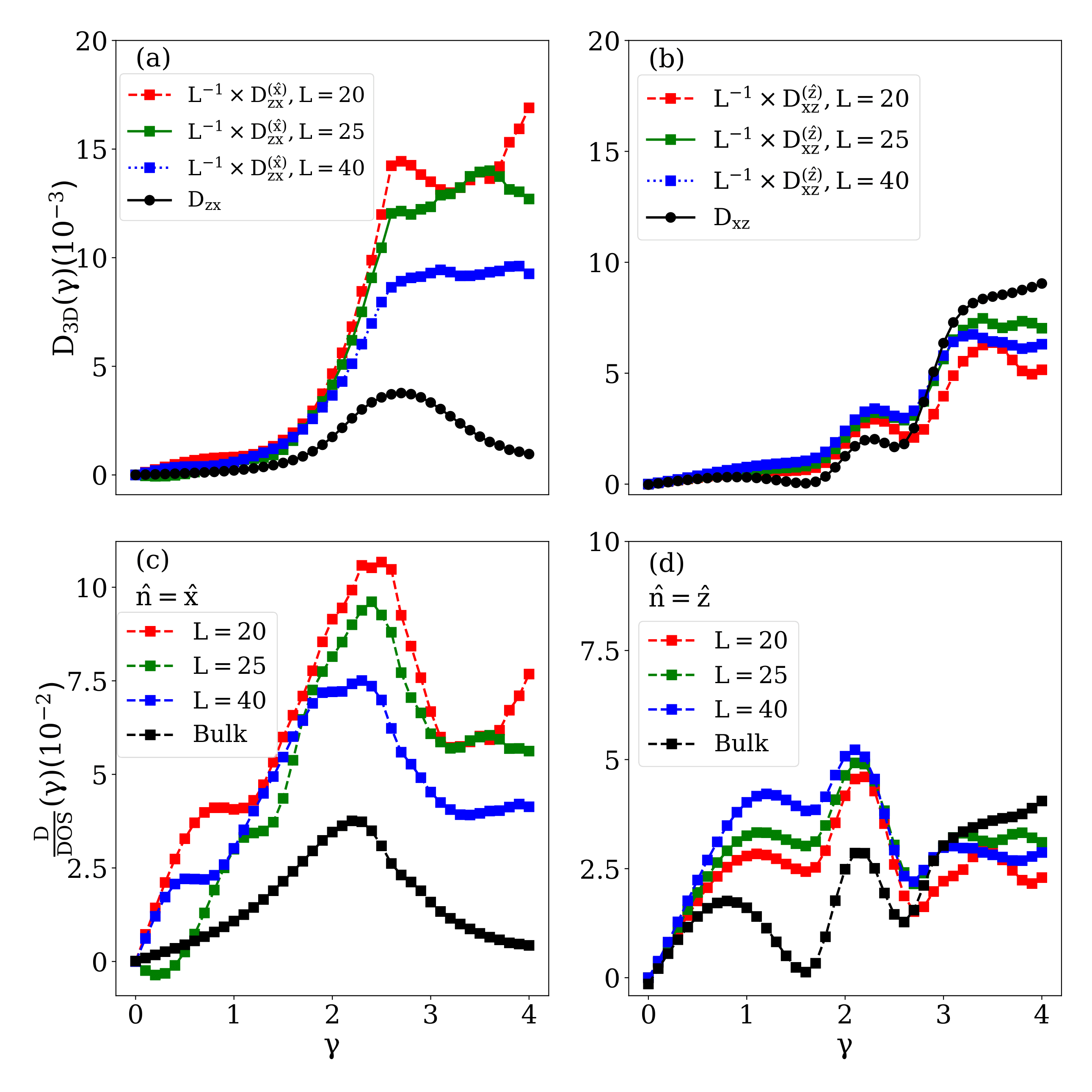}
\caption{(Color online) BCD components as a function of the tilting parameter $\gamma$ for slab and bulk systems. (a) Slab ($D_{zx}^{\hat{x}}/L$) and bulk ($D_{zx}^{\rm 3D}$) components for the $\hat{x}$-cut, for different slab thicknesses. (b) Slab ($D_{xz}^{\hat{z}}/L$) and bulk ($D_{xz}^{\rm 3D}$) components for the $\hat{z}$-cut, for different slab thicknesses.}
\label{FigureTotal}
\end{figure}

Let us now turn our attention towards the NLHE response for the two slabs. We compute the BCD components given by Eq. \eqref{4} in both slab geometries, (a) $D_{zx}^{\hat{x}}/L$ ($\hat{x}$-cut) and (b) $D_{xz}^{\hat{z}}/L$ ($\hat{y}$-cut), as well as in the 3D bulk structure, $D_{zx(xz)}^{\rm 3D}$ (black symbols), in order to better identify the impact of the surface states. Notice that the BCD calculated in the slab geometry is normalized by the number of layers $L$ to allow for a quantitative comparison with the BCD calculated in the bulk. By definition, the BCD is therefore unitless. In the following, we set the chemical potential to $\mu=0.2$ and represent the BCDs as a function of the tilting parameter $\gamma$, as displayed in Fig. \ref{FigureTotal}. In addition, the corresponding ratio between the BCD and the density of states is reported on (c) and (d) and discussed further below.

For the $\hat{x}$-cut [Fig. \ref{FigureTotal}(a)], the bulk BCD ($D_{zx}^{\rm 3D}$, black symbols) displays a nonlinear dependence as a function of the tilting parameter, as observed by Zeng et al. \cite{Zeng}: it first increases smoothly with $\gamma$, reaches a maximum and decreases for large $\gamma$. We attribute this tendency to the contribution of the Fermi surface between the Weyl nodes, as also mentioned by Zeng et al. \cite{Zeng}, and the upper limit that should reach the tilting parameter regarding the inclination of the Weyl nodes. In contrast, the slab BCD ($D_{zx}^{\hat{x}}$, colored symbols) increases sharply and reaches a plateau at large $\gamma$. Furthermore, the value of this plateau is substantially larger than the maximum value obtained in the bulk although it slightly decreases when increasing the number of layers, suggesting that surface states play a major role in the NLHE response. Since the transition between type I and type II regimes produces a change in the configuration of the surface states, with the apparition of Fermi pockets Fig. \ref{Figuredosx}(b)], it is clear that the mismatch between the slab and bulk responses arise from the influence of the Fermi pockets' projections on the BCD across the slab. 

Conversely, for the $\hat{z}$-cut [Fig. \ref{FigureTotal}(b)], the bulk and slab BCDs show a similar behavior, increasing continuously and displaying local plateaus at the type I-type II transition, as well as for large tilting parameter $\gamma$. Based on Fig. \ref{Figuredosz}(b), we infer that the surface track states do not affect qualitatively the behavior of BCD in finite slab samples. In fact, in the $\hat{z}$-cut, the slab BCD converges faster towards the bulk BCD when increasing the number of layers than in the $\hat{x}$-cut configuration. 

Since the density of states of carriers also changes when tuning the band structure, we report the corresponding ratio between the BCD and the density of states in Figs. \ref{FigureTotal}(c) and (d). In Fig. \ref{FigureTotal}(c), for a slab normal to $\hat{x}$, the renormalized BCD exhibits a qualitatively similar trend across the type I-to-type II transition, although the slab BCD is markedly larger than the bulk BCD when approaching the type II regime. Therefore, the Fermi pockets' projections on this slab enhance the value of the BCD due to the reconfiguration of the surface states and the presence of more local states within the unit cell. In contrast, for a slab normal to $\hat{z}$ Fig. \ref{FigureTotal}(d) shows a smaller discrepancy between the slab and bulk calculations suggesting that the BCD is weakly impacted by the track states.

\begin{figure}[ht!]
\includegraphics[width=1\linewidth]{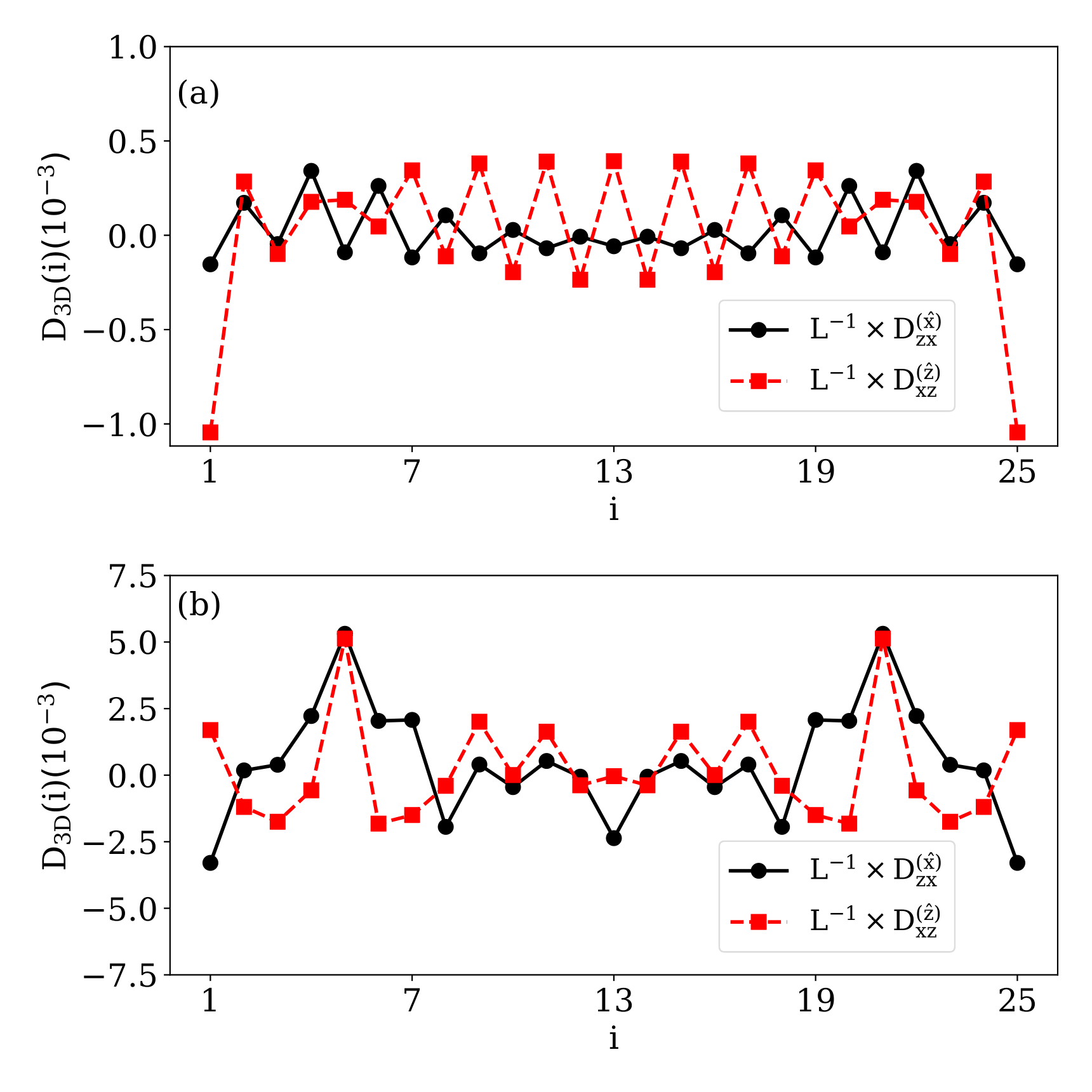}
\caption{(Color online) Layer dependent contribution for the BCD normalized by the number of layers (in this case, $L=25$). We show the tomography for a type I WSM with $\gamma=1$ in (a) and for a type II WSM with $\gamma=3$ in (b).}
\label{FigureTom}
\end{figure}
To complement this analysis, we compute the layer decomposition of the NLHE in slab geometries, for $\gamma=1,\;3$ and $L=25$. The result is reported in Fig. \ref{FigureTom} for (a) $\gamma=1$ and (b) $\gamma=3$ for $\hat{x}$- (black) and $\hat{z}$-cut (red). As we can see from Fig. \ref{FigureTom}(a), in a type I WSM the NLHE response strongly differs at the edges, i.e, at the top ($l$=1) and bottom ($l$=25) layers. From Figs. \ref{Figuredosx}(a) and \ref{Figuredosz}(a) we attribute this behavior the surface states driven by the degenerate Fermi arcs that only appear in the $\hat{x}$-cut. In the case of the type II WSM, Fig. \ref{FigureTom}(b), we note that the magnitude of the central layer ($l=13$) in the $\hat{x}$-cut is larger than that in the $\hat{z}$-cut. Actually, the main contributions stem from layers located underneath the surfaces ($l=5$, $l=21$) rather than on the surface layers. This fact corroborates our claim that in type II WSMs a decisive factor to enhance NLHE is the presence of Fermi pockets' projections at the surface rather than track states.  

\section{Nonlinear Hall effect in WTe$_2$ slabs} \label{IV}

\begin{figure}[ht!]
\includegraphics[width=\linewidth]{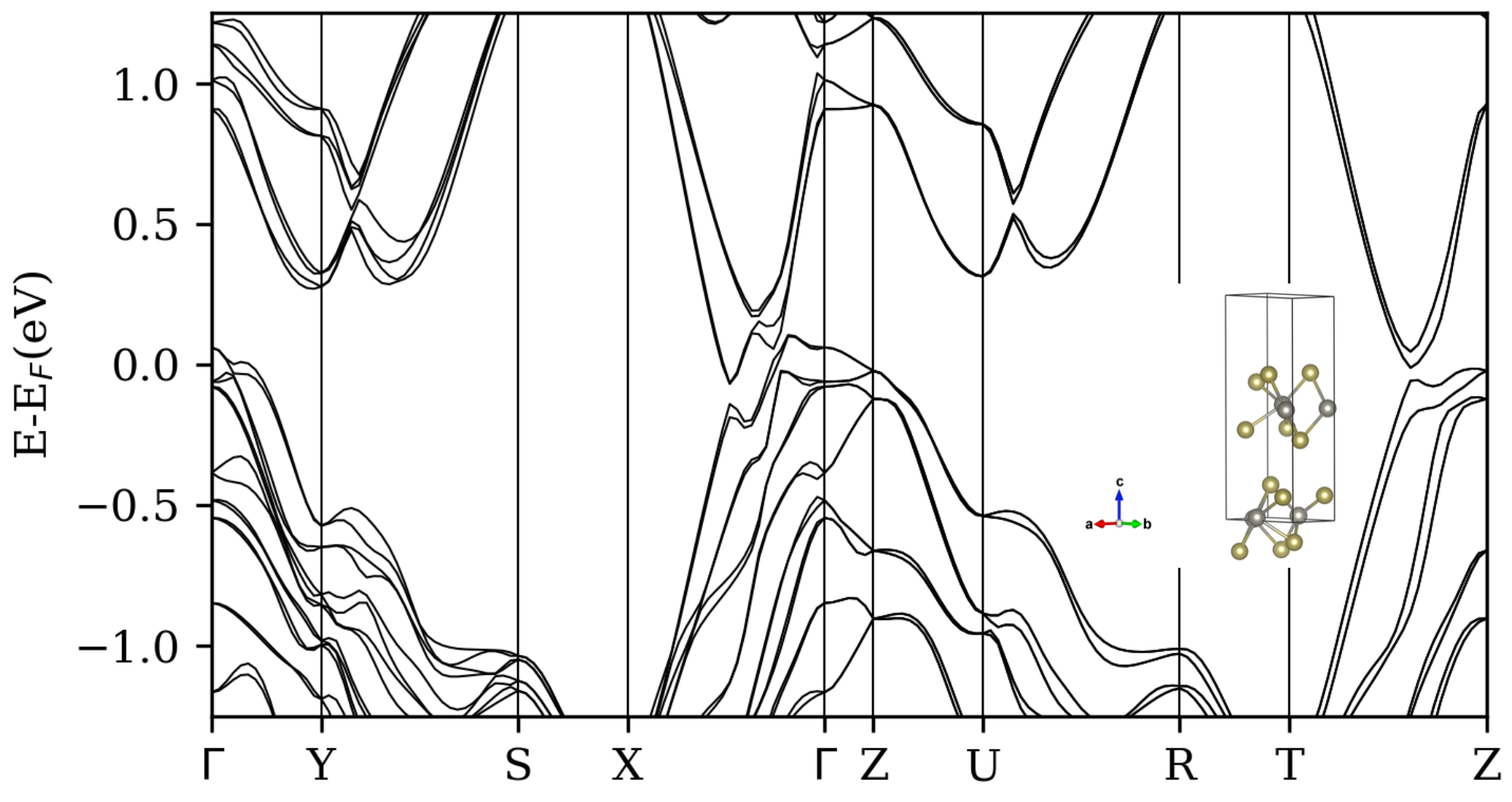}
\caption{(Color online) Bulk band structure of WTe$_2$ type II WSM obtained from density functional theory simulations. The inset displays the unit cell.}
\label{Figure_wte2_bulk}
\end{figure}

Let us now consider a realistic system, WTe$_2$ in its orthorhombic phase, and compute the NLHE from first principles. WTe$_2$ is a well-known type II WSM \cite{Bruno,Wu}, in which NLHE has been originally reported \cite{Ma,Kang}. For the density functional theory simulations \cite{dft1964,dft1965}, we used the Perdew-Burke-Ernzerhof \cite{gga,pbe} exchange-correlation functional. The geometry optimizations were performed using a plane-wave basis as implemented in the Vienna $\textit{Ab-initio}$ Simulation Package (VASP) \cite{vasp1,vasp2}. We have employed 400 eV for the plane-wave expansion cutoff with a force criterion of 5 $\mu$ev/\AA and a reciprocal space sampling containing $16 \times 16 \times 14$ $\mathbf{k}$-points within the Brillouin zone. The ionic potentials were described using the projector augmented-wave (PAW) method \cite{paw}, post-processing calculations were performed using WannierTools \cite{WU2017}. The band structure is displayed in Fig. \ref{Figure_wte2_bulk} with the inset showing the unit cell. The band crossings are located within the ${\rm X}-\Gamma$ path in momentum space, such that the Fermi level was set to zero near this region. With this consideration, we can project the Weyl points on selected surfaces, i.e., $\hat{z}$ (corresponding to the (001) direction) and $\hat{x}$ (corresponding to the (100) direction).

\noindent

\begin{figure}[ht!]
\includegraphics[width=1\linewidth]{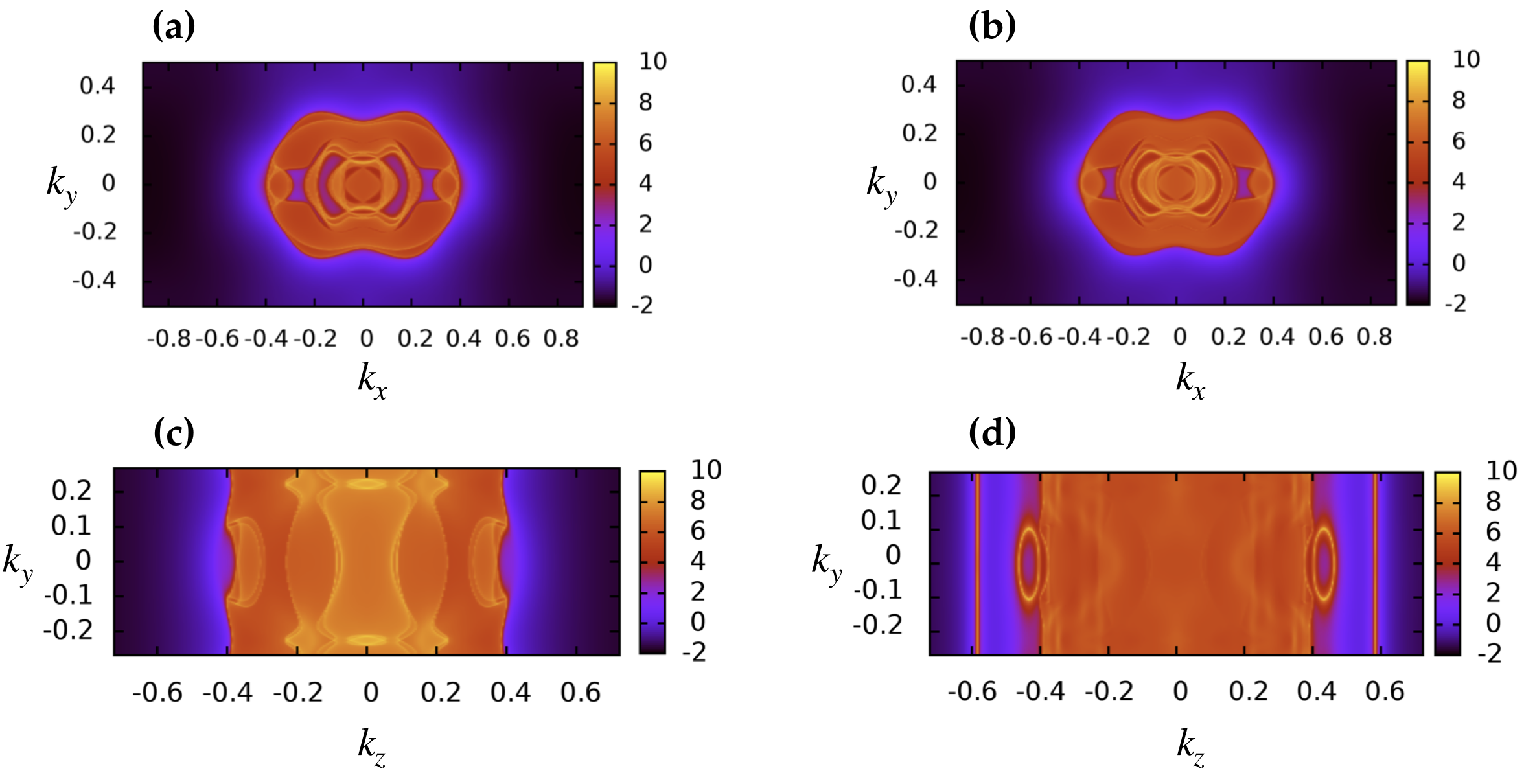}
\caption{(Color online) WTe$_2$ density of states at Fermi level, projected on the bulk (a,c) and top surfaces (b,d) for a slab geometry containing 25 layers. Panels (a,b) correspond to a cut along $\hat{z}$ whereas panels (c,d) correspond to a cut along $\hat{x}$.}
\label{Figure_wte2_surfaces}
\end{figure}

Let us first look at the density of states in the bulk and at the surface. Fig. \ref{Figure_wte2_surfaces} displays the projected density of states in the bulk (a,c) and at the surface (b,d) for a slab cut along the $\hat{z}$ direction (a,b), and for a slab cut along the $\hat{x}$ direction (c,d). As can be observed from Figs. \ref{Figure_wte2_surfaces}(a,b), in the $\hat{z}$-cut the resulting density of states of the slab is larger at the surface than in the bulk, especially near the origin ($\bar{\Gamma}$ point). On the other hand, the opposite situation happens when the slab is normal to $\hat{x}$, as can be noticed from Figs. \ref{Figure_wte2_surfaces}(c,d). From these results and comparing to the lattice model in Figs. \ref{Figuredosx} and \ref{Figuredosz}, we deduce that the surface states highlighted in Fig. \ref{Figure_wte2_surfaces}(b) should enhance the value of the BCD for the slab normal to $\hat{z}$, being strongly dependent on the thickness of the slab, whereas the geometry normal to $\hat{x}$ should be less sensitive to the Fermi arc diversity on the sample.


\begin{figure}[ht!]
\includegraphics[width=\linewidth]{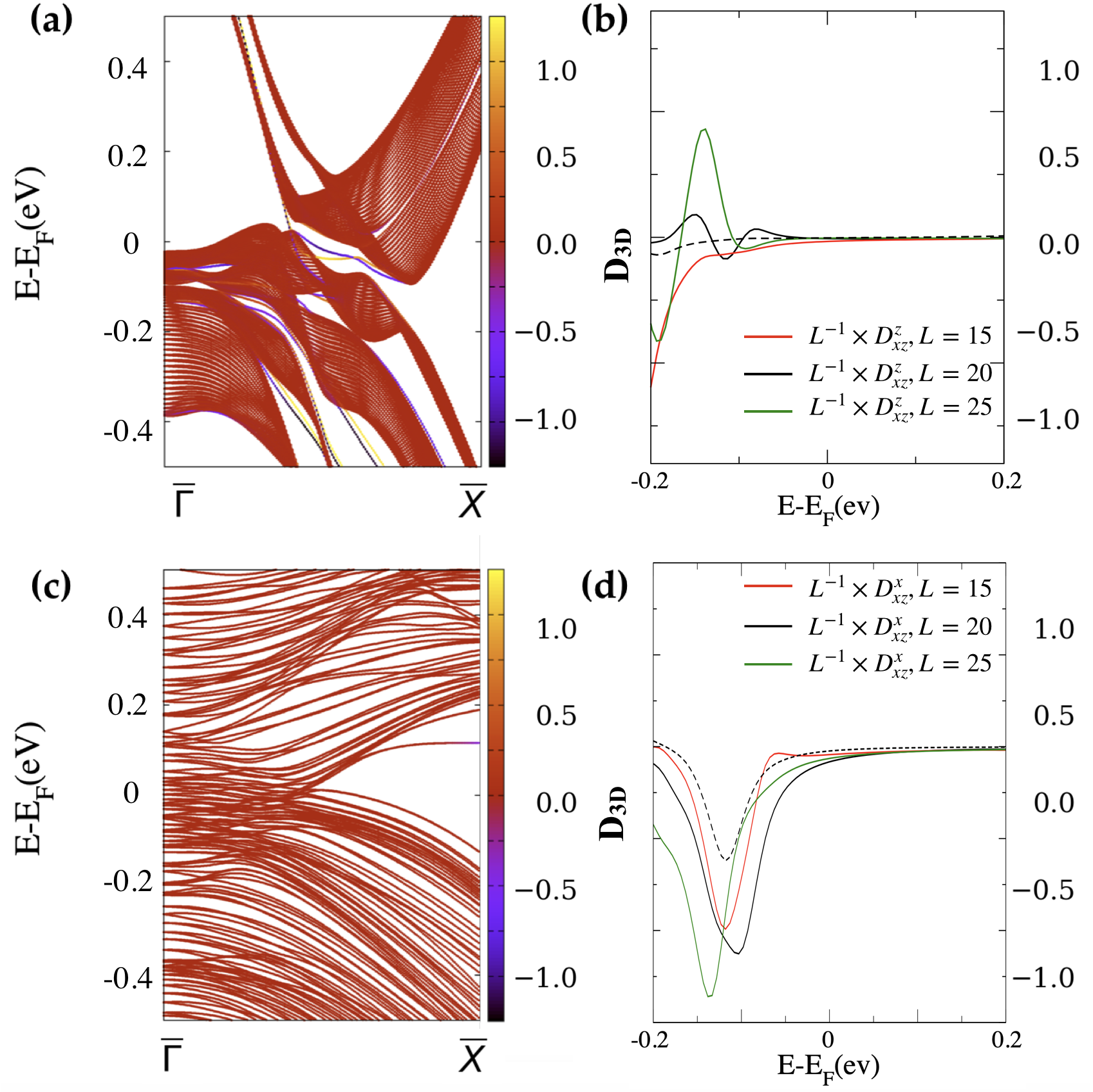}
\caption{(Color online) Band structure of a WTe$_2$ slab containing 25 layers and cut along $\hat{z}$ (a) and $\hat{x}$ (c). For this case the color bar represents the projections on the bottom (-1) and top (+1) layers. (b,d) Corresponding BCD for bulk and slabs containing 15, 20 and 25 layers.}
\label{Figure_wte2_dipole}
\end{figure}

We now move on to the computation of the surface contribution of the NLHE for the two cuts considered above. For the calculation of the BCD, we have used Eq. \eqref{3} performing the sum over a sample Brillouin zone containing $500 \times 500$ k-points. Figures \ref{Figure_wte2_dipole}(a) and (b) display the band structure and the BCD, respectively, computed for a slab cut along the $\hat{z}$ direction. The color bar in (a) represents the projections on the top and bottom surfaces of a slab containing 25 layers. One can distinguish both surfaces by the dark and yellow lines near the Fermi level. In panel (b), the BCD is computed as a function of the energy for slabs containing an increasing number of layers, from 15 to 25. It is clear that the energy profile of the BCD strikingly depends on the slab thickness. This thickness dependence reflects the influence of the surface states. For the sake of comparison, we also reported the value of the BCD computed in the bulk (dashed line). For an infinitely thick slab, the peaks present below the Fermi level and associated with the surface states disappear. Similarly, Figs. \ref{Figure_wte2_dipole}(c,d) display the band structure and BCD, respectively, calculated for a slab cut along the $\hat{x}$ direction. In contrast to the $\hat{z}$-cut discussed above, the surface states cannot be clearly identified in the band structure that is instead dominated by bulk states. In panel (d), the BCD is computed for three different slab thicknesses. Interestingly, the qualitative behavior of the BCD is similar, displaying a peak whose position is weakly influenced by the slab thickness. For an infinitely thick slab, the bulk BCD (dashed line) conserves the overall structure, with a slight reduction in magnitude. These calculations show that surface states do substantially impact the NLHE in realistic materials.

\section{Conclusions} \label{V}

We have investigated the influence of surface states on the NLHE response of non-centrosymmetric time-reversal invariant WSMs. Using both a model Hamiltonian and realistic first principles calculations, we have demonstrated that depending on the direction of the cut, surface states emerge that can substantially contribute to the NLHE of the slab. Notice that the topological nature of the surface state (topologically protected arcs, or track states) has no impact on the overall BCD, only the relative number of states occupying the surface and the bulk matters. We emphasize that the relative contribution of the surface states with respect to the bulk states is in fact very large, leading to dramatic thickness-dependence of the NLHE response, in particular in type-II WSMs. This observation, confirmed by first principles calculations on WTe$_2$ slab geometries, suggests that surface states can contribute much more efficiently to NLHE than their bulk counterpart. These conclusions are not limited to WSMs and should apply to other topological systems and topologically trivial classes of non-centrosymmetric materials.

\begin{acknowledgments}
D.G.O thanks to Andr\'es Sa\'ul and Carlos Quezada for their help and fruitful discussions related to the computational implementation of the work. Besides, D.G.O and A.M acknowledge support from the Excellence Initiative of Aix-Marseille Universit\'e - A*Midex, a French ”Investissements d’Avenir” program. 
\end{acknowledgments}

\bibliography{WTom_Ref}
\end{document}